\documentclass[iop, apjl, twocolappendix, twocolumn]{aastex631}
\usepackage{graphicx}% Include figure files
\usepackage{bm}% bold math
\usepackage{amsmath,amssymb}
\usepackage{natbib}
\usepackage{booktabs}
\usepackage{color}
\usepackage{units}

\usepackage[normalem]{ulem}

\newcommand{\program}[1]{\textsc{#1}}
\newcommand{\citeg}[1]{\citep[e.g.,][]{#1}}

\newcommand{\be}{\begin{equation}}
\newcommand{\ee}{\end{equation}}

\begin{document}
\title{Cautionary tales on heating-rate prescriptions in kilonovae}

\author[0000-0003-2700-1030]{Nikhil Sarin}
\affil{Oskar Klein Centre for Cosmoparticle Physics, Department of Physics,
Stockholm University, AlbaNova, Stockholm SE-106 91, Sweden}
\affil{Nordita, Stockholm University and KTH Royal Institute of Technology \\
Hannes Alfvéns väg 12, SE-106 91 Stockholm, Sweden}
\author[0000-0002-3833-8520]{Stephan Rosswog}
\affil{University of Hamburg, Hamburger Sternwarte, Gojenbergsweg 112, 21029, Hamburg, Germany}
\affil{The Oskar Klein Centre, Department of Astronomy, AlbaNova, Stockholm University, SE-106 91 Stockholm, Sweden}

\begin{abstract}
A major ingredient for kilonova lightcurves is the radioactive heating rate and its dependence on the electron fraction and velocity of the ejecta and, in principle, on the nuclear mass formula. 
Heating-rate formulae commonly used as the basis for kilonova models previously employed in the literature produce substantially different outputs for high electron fractions ($Y_{e} \gtrsim 0.3$) and at late times ($t \gtrsim \unit[1]{d}$) compared to newer prescriptions.
Here, we employ standard semi-analytical models for kilonovae with better heating rate prescriptions valid for the full parameter space of kilonova velocities and electron fractions to explore the impact of the heating rate on kilonova lightcurves. 
We show the dangers of using inappropriate heating rate estimates by simulating realistic observations and inferring the kilonova parameters via a misspecified heating-rate prescription. 
While providing great fits to the photometry, an incorrect heating-rate prescription fails to recover the input ejecta masses with a bias significantly larger than the typical statistical uncertainty. 
This bias from an incorrect prescription has significant consequences for interpreting kilonovae, their use as additional components in gamma-ray burst afterglows, and understanding their role in cosmic chemical evolution or for multi-messenger constraints on the nuclear equation of state. We showcase a framework and tool to better determine the impact of different modelling assumptions and uncertainties on inferences into kilonova properties.
\end{abstract}

\section{Introduction}\label{sec:intro}
The electromagnetic observations of AT2017gfo~\citep{smartt17, villar17, cowperthwaite17, tanvir17, kasen17, evans17}, the kilonova that accompanied the first gravitational-wave observation of a binary neutron star merger~\citep{abbott17b, abbott17c}, confirmed that binary neutron star mergers are sites of rapid neutron capture (r-process) nucleosynthesis as theoretically expected~\citep{lattimer74,symbalisty82,eichler89,rosswog99,freiburghaus99b}. Detailed spectroscopic observations and modelling would later suggest the production of specific r-process elements such as Strontium~\citep{watson19}. 
The combination of electromagnetic and gravitational-wave observations of GW170817 would go on to provide many constraints on significant questions in physics and astrophysics, such as a standard-siren measurement of the Hubble constant, independent of the cosmic distance ladder~\citep{abbott17a}, constraints on the nuclear equation of state~\citep{margalit17, Coughlin2019, Nicholl2021}, and 
the confirmation that gravitational waves propagate to an enormous accuracy at the speed of light ~\citep{abbott17d}. 

Since the groundbreaking discovery of AT2017gfo, kilonova observations have been lacking. Curiously, our two best kilonova candidates are also ones from an unexpected source: two \textit{long-duration} gamma-ray bursts, GRB 211211A~\citep{rastinejad22} and GRB230307A~\citep{Levan2024}, with the kilonova-like excess seen in the afterglows of both GRBs, similar to how supernovae emerge as bumps in the afterglows of long gamma-ray bursts~\citep{galama98,Cano2017}. 

\begin{figure*}[t]
    \centering
    \includegraphics[width=0.95\textwidth]{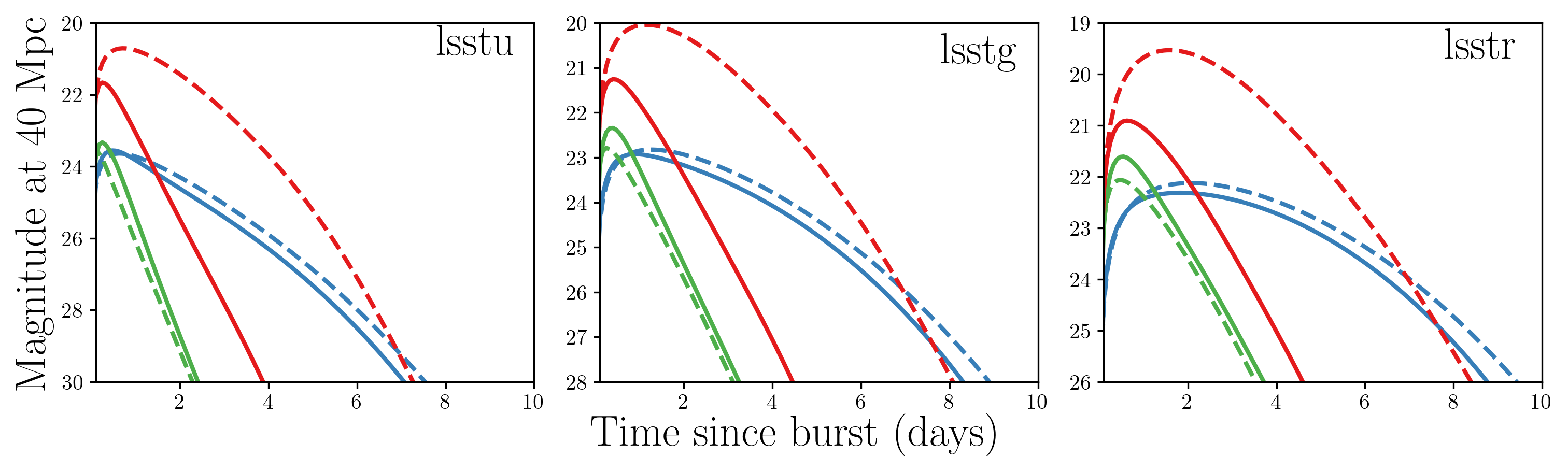}
    \caption{Set of lightcurves for different Vera Rubin Observatory filters with a two-component kilonova model for different ejecta velocities and electron fractions (other parameters are identical). For all three models we fix the ejecta masses of each component to $0.02$ and $0.04~M_{\odot}$, and floor temperatures to $1500$ and $500$ K, respectively to isolate the impact of the heating rate. The blue and green curves show two component ejecta's with electron fractions of $0.35$, $0.05$, and $0.25$ and $0.1$, while the red curves show both components for an electron fraction of $0.4$. For the ejecta velocities, we fix the first component velocity to $0.05~c$ for the blue and red curves and $0.25~c$ for the green curve. The second component velocities are $0.1$, $0.05$, and $0.15~c$, for the blue, green, and red curves, respectively. The old heating-rate prescription~\citep{korobkin12a} is shown as dashed curves while the solid curves show the updated prescription~\citep{rosswog24a}. The physics behind the two models is the same apart from the heating-rate prescription.}
    \label{fig:opticallcs}
\end{figure*}

The detection and interpretation of kilonovae light curves are heavily reliant on available kilonova models. Specifically, such models form a key aspect for designing surveys, understanding depth requirements for follow-up, and ultimately characterising and inferring the properties of candidate events. 
However, kilonova light curve models are subject to a number of different ingredients, each with its own underlying uncertainty. 
One of these critical modelling ingredients is the nuclear heating rate, i.e., the energy produced per second by one gram of r-process material, which is ultimately subject to the decay channels of r-process nuclei~\citep[see e.g.,][for an extensive review on various aspects of rapid neutron capture nucleosynthesis]{cowan21}. We note that the term ``heating rate" is often used to describe the combination of the ``naked'' heating rate and a thermalization efficiency. In this work, we use heating rate to refer to the ``naked'' heating rate.

The temporal decay of a kilonova lightcurve has significant diagnostic power and may provide clues to the ejecta composition. 
If a broad distribution of nuclei is available and all are decaying exponentially, one expects a power-law dependence of the heating rate as a function of time~\citep{way48, li98, metzger10b,hotokezaka17a}. 
If, instead, only a few isotopes dominate the heating rate, one expects distinctive ``bumps" in the lightcurve that can provide constraints on the half-lives of the decaying species~\citep{grossman14a,martin15,lippuner15,rosswog18a,wanajo18,wu17,kasliwal22}.

The specific decay channels of radionuclides that occur and their decay products determine how much energy is deposited in the expanding ejecta cloud and the overall heating rate. Any neutrino products escape the ejecta and do not contribute to the overall heating rate, while the heating-rate efficiency of $\gamma-$rays decays very rapidly and becomes negligible after tens of days \citep{hotokezaka16a}. 
On the other hand, charged particles, such as $\alpha$-particles and fission products, keep depositing energy into the ejecta even at late times, and this makes kilonova lightcurves sensitive to the contribution of those decay channels~\citep{barnes16a,rosswog17a,kasen19,wu19,zhu18,vassh19,giuliani20}. 

Understanding what decay channels take place and what their decay products are is non-trivial. To complicate matters further, the interactions themselves are sensitive to the overall composition of the kilonova ejecta, which itself is linked in a multitude of ways to the progenitor binary neutron star system, equation of state, and the fate of the remnant~\citeg{rosswog18a, Sarin_review}. 
For relatively high $Y_e$ values, $\beta$-decays dominate, and only the resulting electrons and photons are relevant for heating the ejecta. 
In low $Y_e$ ejecta, actinides can be produced which, via $\alpha$-decays and fission, can deposit large amounts of energy, altering the heating rate by up to an order of magnitude depending on the underlying nuclear mass model \citep{barnes16a,rosswog17a,wu19, Zhu2021, Barnes2021, Lund2023}.

In principle, kilonova models should be calculated self-consistently, such that feedback from the energy released and how this affects the expansion of the ejecta and nucleosynthesis/heating rate is incorporated. Recent work suggests that failure to include this feedback can compromise the abundance evolution~\citep{magistrelli24}. Moreover, kilonova models should keep track of the exact decay channels to determine both the ``naked'' heating rate (i.e., the raw output) and the thermalization efficiency (the fraction that is coupled to the ejecta). 
This is difficult in practice since such models would need to be coupled to nuclear reaction networks which is computationally prohobitive.
Instead, it is common to construct separate expressions for the naked heating rate, $\dot{q}$, and the thermalization efficiency, $\epsilon_{\rm th}$. The effective heating rate is then $\dot{q}_{\rm eff}= \epsilon_{\rm th} \; \dot{q}$.

A popular heating-rate prescription is~\citep{korobkin12a}
\be\label{eq:old_prescription}
% \dot{q}_{\rm eff}= \dot{q}_0 \left( \frac{1}{2} - \frac{1}{\pi} {\rm arctan} \frac{t-t_0}{\sigma}\right)^\alpha \times \left( \frac{f}{0.5}\right),
\dot{q}= 2q_0 \left( \frac{1}{2} - \frac{1}{\pi} {\rm arctan} \frac{t-t_0}{\sigma}\right)^\alpha,
\ee
with $q_0 = 2 \times 10^{18}$ erg/(g s), $t_0=1.3$ s, $\sigma= 0.11$ s, $\alpha= 1.3$~\citep{korobkin12a}, consistent with heating-rate estimates from \citet{metzger10a}, and the thermalization efficiency, $\epsilon_{\rm th}$ computed following~\citet{barnes16a} or~\citet{wollaeger18a}. 
This specific heating-rate prescription was obtained as a fit to the heating rate from a broad set of Newtonian+GW-backreaction hydrodynamics simulations with the stiff equation of state of \cite{shen98a,shen98b} and a multi-flavour leakage scheme \citep{rosswog03a}. This prescription has served as the basis for a large variety of kilonova models, including the majority of model fits to the lightcurve of AT2017gfo, GRB211211A, and GRB230307A~\citep{villar17, Nicholl2021, rastinejad22, Levan2024}, and detailed radiative transfer simulations~\citep{bulla19a}.

\begin{figure*}[t]
    \centering
    \includegraphics[width=1.0\textwidth]{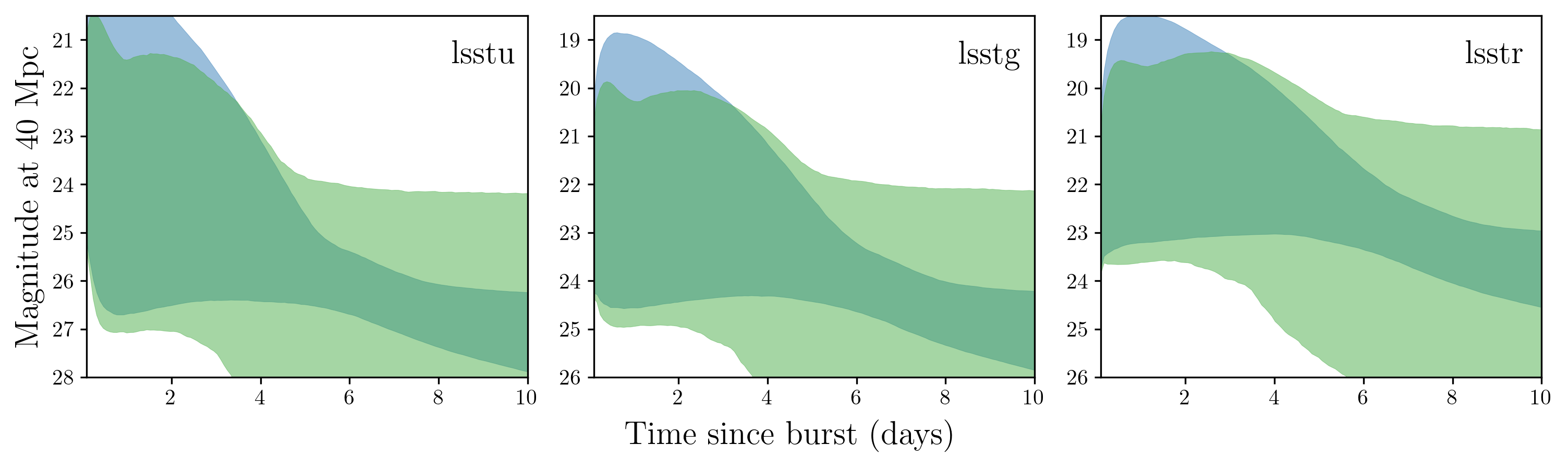}
    \caption{$95\%$ credible interval from $5000$ simulated kilonovae with fixed ejecta masses and temperature floors and varying velocities and opacities for old prescription in blue and updated prescription in green.}
    \label{fig:poplcs}
\end{figure*}

While the leakage scheme used in these simulations includes positron captures and, therefore, allows for an increase in the electron fraction, it does not account for neutrino absorption and both the essentially Newtonian gravity and the stiff equation of state favour only moderate temperatures during the merger. 
As a result, the ejecta was strongly dominated by ``tidal ejecta" at the original neutron star electron fraction, $Y_e\approx 0.04$, i.e., the heating-rate fit was based on simulations for $Y_{e} \approx 0.04$, which is inappropriate for shock- and wind-ejecta. 

Full general-relativistic simulations with more sophisticated neutrino treatments, produce, as expected, a broader range of electron fractions~\citep{perego19a,shibata19a}, which are not adequately captured by the previous prescription~\citep{rosswog24a}. 
The fit in the latter reference is based on a library of parametrized trajectories where velocity and electron fraction were systematically varied. The actual nuclear network calculations were performed with WinNet \citep{winteler12,winteler12b,reichert23} which is based on the BasNet network \citep{thielemann11}. The functional dependence of the fit is more complicated than the above prescription, with multiple additional terms and written as, 
\begin{equation}\label{eq:new_prescription}
\begin{array}{r}
\dot{q}=q_0\left(\frac{1}{2}-\frac{1}{\pi} \arctan \frac{t-t_0}{\sigma}\right)^\alpha\left(\frac{1}{2}+\frac{1}{\pi} \arctan \frac{t-t_1}{\sigma_1}\right)^{\alpha_1}+ \\
+C_1 e^{-t / \tau_1}+C_2 e^{-t / \tau_2}+C_3 e^{-t / \tau_3},
\end{array}
\end{equation}
where the coefficients no longer have fixed values, but depend on the velocity and electron fraction of the ejecta. We refer to the appendix of \cite{rosswog24a} for the explicit coefficient expressions\footnote{We note that successfully implementing this prescription and interpolating to different values of $Y_{e}$ and ejecta velocity requires the coefficients to be cast in a more numerically stable form and the interpolant to be built on $\dot{q}$ directly instead of using the functional form with interpolated coefficient terms. We refer the interested reader to the {\sc Redback} implementation on GitHub \url{https://github.com/nikhil-sarin/redback}}. 
The role of this updated heating-rate prescription has been explored in detailed radiative transfer simulations of kilonovae~\citep{Bulla2023}, or used for predictions for surveys like LSST using the Vera Rubin Observatory~\citep{Setzer2023}. However, it was not explored in detail for semi-analytical models. 
Moreover, it is unclear what the consequences of using the previous heating-rate prescription in fitting light curves are and what the systematic uncertainty for inferences of kilonova ejecta parameters due to uncertainties in the heating rate truly is.

As discussed in \citet{rosswog24a}, the updated prescription produces substantially different heating rates for $Y_{e} \gtrsim 0.3$ and at times ($t \gtrsim \unit[1]{d}$). Therefore, for the majority of this \emph{Letter}, we will focus more on the impact of the updated heating rate prescription of~\citet{rosswog24a} in this regime.
In particular, in Sec.~\ref{sec:models}, we briefly describe a semi-analytical kilonova model implemented in open-source software, \program{Redback}~\citep{sarin23_redback}, highlighting the differences in individual light curves and a population due to the heating-rate prescription alone. 
We follow this by showcasing the dangers of using the previous heating-rate prescription to fit realistic, simulated kilonovae observations in Sec.~\ref{sec:misspecify}. 
We then discuss the implications of uncertainty in the heating-rate prescriptions and conclude in Secs.~\ref{sec:implications} and \ref{sec:discussion}, respectively. 
%%%%%%%%%%%%%%%%%%%%%%%%%%%%
\section{Modelling and impact of heating-rate}\label{sec:models}
Early modelling of the photometry of AT2017gfo made it abundantly clear that one single ejecta composition was inadequate for explaining the observations, with a clear evolution from a bright blue to a red source~\citep{cowperthwaite17, kasen17, metzger19a}. 
This led to the development of multi-component models, where each distinct ejecta component is characterized by a single opacity~\citep{kasen17}. 
For the specific case of AT2017gfo, a two-component model with one lanthanide-poor, blue component and another lanthanide-rich red component proved sufficient to provide adequate fits to the observations~\citep{villar17}, with extensions to include the effects of other emission mechanisms such as shock-cooling~\citep{arcavi17}, effects of the hypermassive neutron star wind~\citep{Metzger2018}, or interaction with a jet~\citep{kasliwal17}, which served to improve fits to the observed data. 

Although these extensions are relevant, the basic ingredients of the majority of semi-analytical models for kilonovae are broadly similar. We first have the `naked' heating-rate prescription, which, for the majority of models follows Eq.~\ref{eq:old_prescription}.
Only some fraction of this input energy is available to thermalize this ejecta, which is often taken from~\citet{barnes16a}, 
\begin{equation}
\epsilon_{\mathrm{th}}(t)=0.36\left[e^{-a t}+\frac{\ln \left(1+2 b t^d\right)}{2 b t^d}\right]. 
\end{equation}
Here, a, b, c are functions of the ejecta velocity and mass, with the overall function built as a fit to simulations~\citep{barnes16a}. 
The overall bolometric luminosity is then estimated following the Arnett model for supernovae~\citep{arnett82}, with a temperature and wavelength independent gray opacity, which is estimated from the electron fraction, $Y_{e}$ following e.g.,~\citet{tanaka20a}. We note that this conversion is also a source of systematic uncertainty, which we will return to in later sections.
As most comparisons to data are done through observations in specific filters, an assumption for the spectral energy distribution is also required, with the typical choice being a blackbody with a floor temperature such that the photospheric radius starts receding and the temperature remains constant once the floor temperature is reached. We note that the opacity and therefore $Y_{e}$, is often fixed to a particular value in several analyses previously performed in the literature~\citep{cowperthwaite17, villar17, Nicholl2021}.

The above describes the ingredients for a basic one-component kilonova. A multi-component model is assumed to be two or more of these components evolved independently with the output a sum of the individual components and often with each individual component at a fixed gray opacity. These simplified assumptions are of course, far from correct. However, for the purposes of interpreting observations and designing surveys such models form a critical building block, which can then be built upon by more expensive and detailed radiative transfer simulations. Furthermore, such models are also typically much quicker to adapt to additional physics compared to expensive and detailed radiative transfer simulations. 

The focus of this work is to highlight the impact of the `naked' heating-rate prescription on kilonova lightcurves and show the impact of using an incorrect prescription and the systematic uncertainty due to our current uncertain knowledge of the true heating rate. Therefore, we limit our modelling to the standard approach as outlined above forgoing more detailed treatments of e.g.,
the thermalization efficiency and its dependence on the electron fraction~\citep{Shenhar2024, Guttman2024}, which are critical to accurately capture kilonova lightcurves and infer ejecta properties from observations.

As we described in the introduction, the heating-rate prescription used by these models is based on simulations where the ejecta was predominantly at an electron fraction $Y_{e} = 0.04$, i.e., (following the conversion from~\citet[][see their Table 1]{tanaka20a}), a grey opacity of $\unit[38]{g^{-1}~cm^{2}}$, in stark contrast to the opacities fixed for different components when fitting AT2017gfo, and other kilonova candidates. For example, the two-component fit for AT2017gfo in~\citet{villar17} had gray opacities of $\unit[0.5]{g^{-1}~cm^{2}}$ and $\unit[10]{g^{-1}~cm^{2}}$. 
We note that as the heating rates are broadly consistent for $Y_{e} \lesssim 0.015$ due to fission cycling, the impact of the opacity is likely small for grey opacities $\gtrsim 30$. 
A different heating-rate prescription has a dramatic impact on the inferred properties of the ejecta (we explore this in more detail in Sec.~\ref{sec:misspecify}). 
However, first we investigate the impact of the updated heating-rate prescription on the kilonova lightcurves themselves. 
Throughout this work, we assume the electron fraction to opacity relationship from \citet{tanaka20a}.

\begin{figure*}[t]
    \centering
    \includegraphics[width=1.0\textwidth]{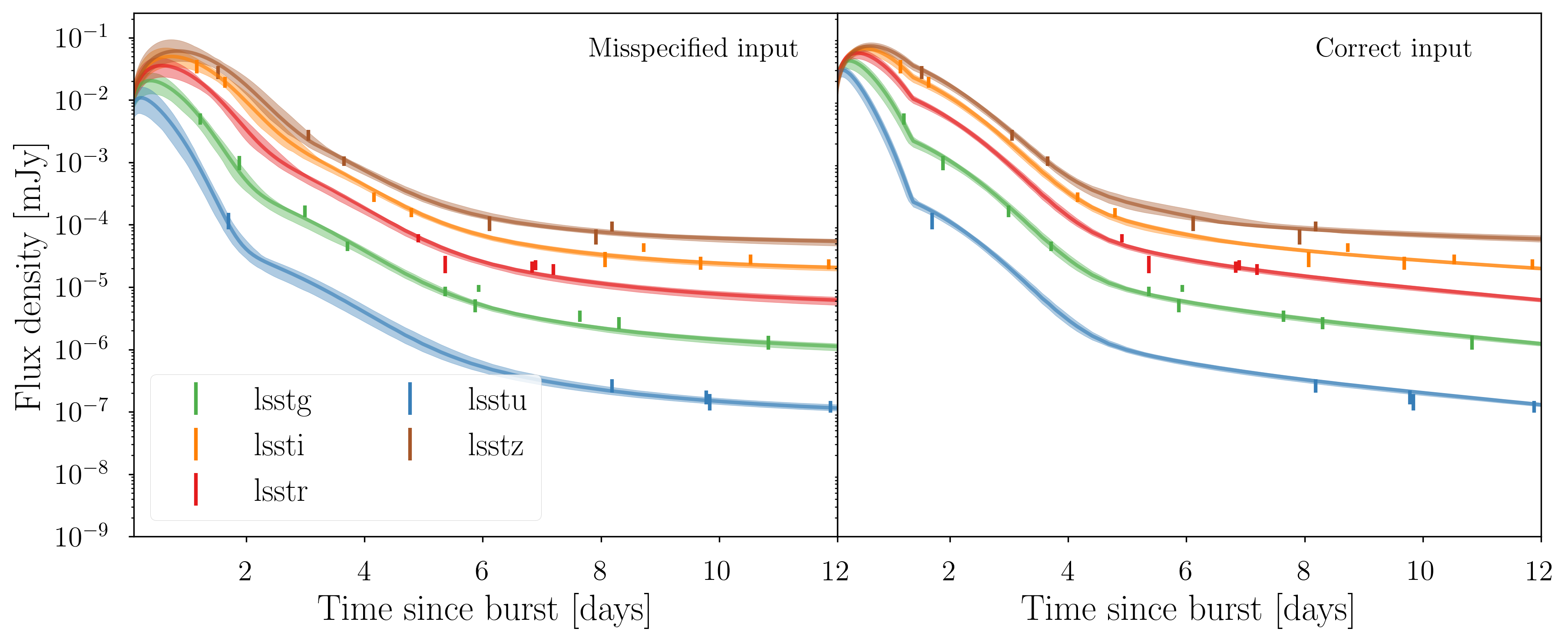}
    \caption{Simulated data from the Vera Rubin Observatory for different bands using a two component kilonova simulated with the updated heating-rate prescription and $95\%$ credible interval fit with old heating rate prescription~\citep{korobkin12a} (left panel) and the updated prescription~\citep{rosswog24a} (right panel).}
    \label{fig:fittedlcs}
\end{figure*}

In Fig.~\ref{fig:opticallcs}, we show three representative two component kilonova model magnitudes at $\unit[40]{Mpc}$. Here the difference is only due to the heating-rate prescription, with the old prescription~\citep{korobkin12a} shown as dashed curves while the solid curve is the updated prescription~\citep{rosswog24a}. For all three models we fix the ejecta masses of each component to $0.02$ and $0.04~M_{\odot}$, and floor temperatures to $1500$ and $500$ K, respectively to isolate the impact of the heating rate. The blue and green curves show two component ejecta with 
electron fractions of $0.35$, $0.05$, and $0.25$ and $0.1$, while the red curves show both components for an electron fraction of $0.4$. 
For the ejecta velocities, we fix the first component velocity to $0.05$c for the blue and red curves, and $0.25~c$ for the green curve, while the second component velocities are $0.1$c, $0.05$c, and $0.15$c, for the blue, green, and red curves, respectively. 
The red curves demonstrate the significant impact of the heating-rate prescription, not only in changing the magnitude at peak by up to $1$ mag, but also the temporal evolution of the lightcurve, in particular the decline rate. 
In general, across a broad range of parameters, we find that the previous prescription produces brighter kilonovae at peak, but the late-time evolution is generally faster, although the parameters for the green curve indicate the new prescription is brighter. 

The above detailed lightcurves highlight specific cases where the updated heating-rate prescription leads to a markedly different lightcurve. However, it is also instructive, to look at the trend from a broader set of lightcurves. 
To understand the impact across a broad sample, we create $5000$ lightcurves drawn from uniform priors on the electron fractions between $0.05$ and $0.5$. 
Truncated normal priors for velocity with the first component with $\mu = 0.15$c, $\sigma = 0.1$c, and the second component with $\mu = 0.1$c, $\sigma = 0.05$c. For both components we fix the minimum and maximum velocities to $0.05$c and $0.35$c, respectively. 
We also fix the ejecta masses and temperature floors to $0.01$, $0.03~M_{\odot}$ and $3500$ and $1500$ K, again to isolate the impact of the heating-rate prescription. In reality, the properties of the ejecta, such as the masses, opacities, velocities are all correlated and link to the properties of the binary, e.g., through relations calibrated to numerical relativity simulations~\citep{Coughlin2019, Setzer2023}. 
However, these relations are largely uncertain and can at times make nonsensical predictions outside the regime of the small subset of numerical simulations they are based on~\citep{Henkel2023}, something we wish to avoid. 

In Fig.~\ref{fig:poplcs}, we show the $95\%$ credible interval from the sample of $5000$ lightcurves, with the old prescription in blue and the updated heating-rate prescription in green. In other words, across our simulated set of lightcurves, at any given time the magnitude of the kilonova at $\unit[40]{Mpc}$ lies within the shaded band for each model at $95\%$ confidence. These lightcurve distributions highlight the broad impact of the updated heating-rate prescription, with the new prescription in general result in a dimmer kilonova at peak but with greater variation in decline rates. Another feature these lightcurve distributions highlight is the significantly tighter shape of the previous prescription (blue) i.e., the evolution of the lightcurve with the previous prescription is largely similar regardless of the opacity or ejecta velocity. This is in stark contrast to the updated prescription, where the lightcurve evolution depends sensitively on the opacity and ejecta velocity, a direct consequence of their impact on the input heating-rate. 
%%%%%%%%%%%%%%%%%%%%%%%%%%%%
\vspace{1cm}
\section{Bias due to misspecified prescription}\label{sec:misspecify}

As we alluded to above, a critical question to investigate is what is the impact of using the incorrect prescription for the heating rate given observational uncertainties. This is pertinent as if the impact is smaller than observational uncertainties, then pursuing a more detailed prescription is largely an academic exercise. The impact shown in the lightcurves in Sec.~\ref{sec:models} suggests a significant difference, here, we aim to explore what the impact translates to in terms of fitting observations. We note while we only apply this methodology here to exploring the impact of differences in ``naked'' heating-rate prescriptions, such a procedure is equally valid to explore the impact of other systematics in kilonova modelling.

We simulate target-of-opportunity style observations with the Vera Rubin Observatory~\citep{Ivezic2019} for a kilonova with the updated heating-rate prescription using \program{Redback}. 
In particular, we set the input parameters for the simulated data to $0.015$ and $0.03~M_{\odot}$ for the ejecta mass of the two components, with floor temperatures of $3000$ and $1200$~K. We set the velocities to $0.25$ and $0.1$c and the electron fractions to $0.4$ and $0.15$.
We then fit the simulated data with a two component kilonova model with the old heating-rate prescription~\citep{korobkin12a}, i.e., a misspecified model and the updated prescription using \program{Redback} with the \program{pymultinest}~\citep{Buchner2016} sampler implemented in \program{bilby}~\citep{Ashton2019a}. We note that we fit for all $8$ model parameters i.e., the ejecta mass, velocity, floor temperature and electron fractions for each component. We also place two constraints on our prior, such that the ejecta mass of the second component (likely the disk-wind) is larger but moving at a slower velocity than the dynamical ejecta, consistent with expectations from numerical simulations~\citeg{perego14b,dietrich17b,siegel18}. 

In Fig.~\ref{fig:fittedlcs}, we show this data and fits with a misspecified model (in the left panel) and the correct model (right panel). These fits highlight a key point, by eye, even an incorrect prescription can provide great fits to the observations, that would not lead to any suspicion that something is wrong with the models. This highlights the need to check for consistencies from inferences on lightcurves with other observables such as overall constraints on energetics or the spectrum.
We note that methods such as Bayesian model selection could elucidate the correct model with a $\ln \rm{BF} \approx 20$, i.e., a $e^{20}$ preference for the updated, (input) heating-rate prescription. However, Bayesian model selection is often not practical outside of simulated data sets where we can control all aspects of data generation.  

\begin{figure}[t]
    \centering
    \includegraphics[width=1.0\columnwidth]{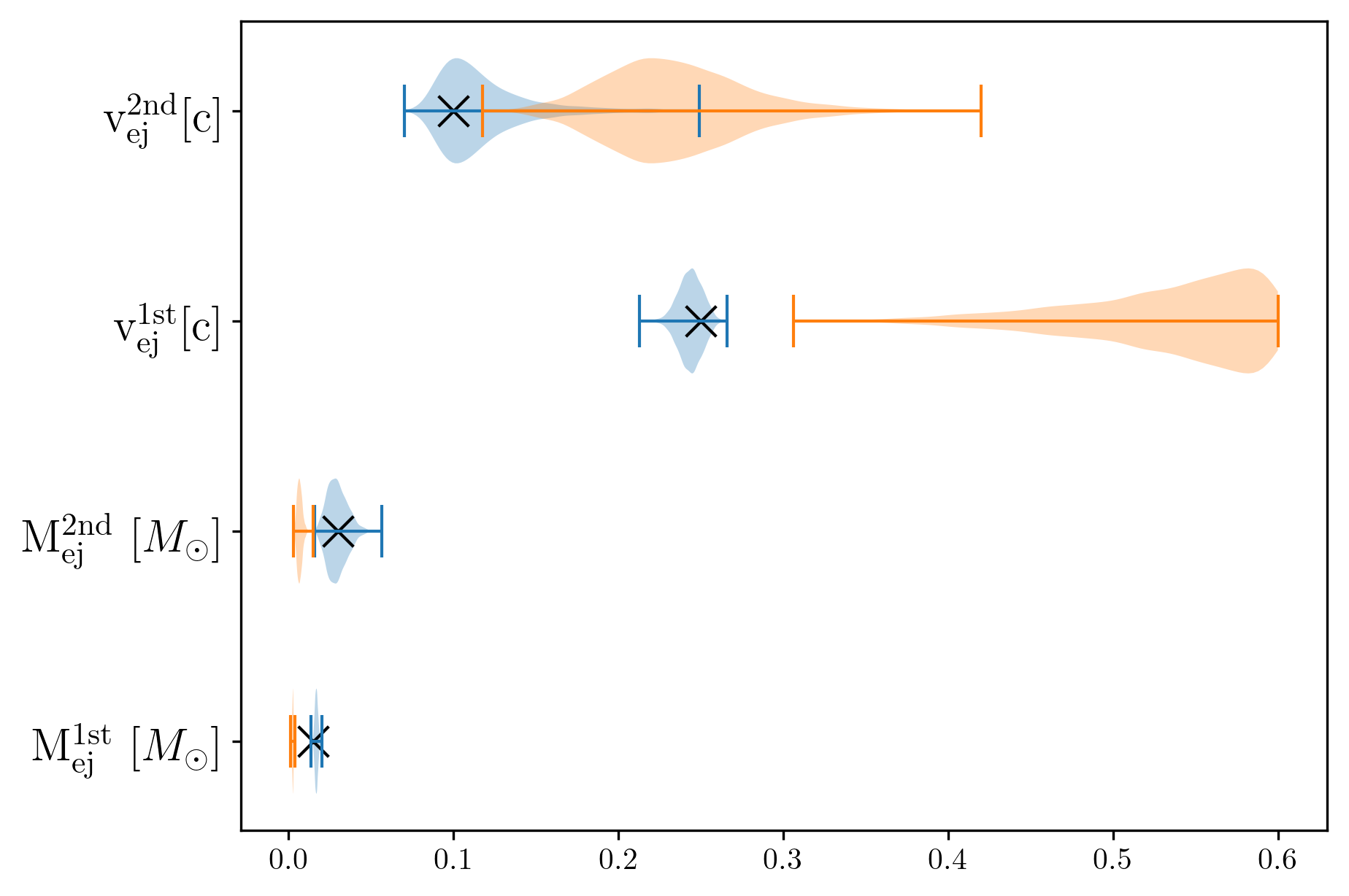}
    \caption{Recovered posteriors from old (orange) and updated (blue) heating-rate prescriptions. The black crosses indicate the true input of the simulated data.}
    \label{fig:posteriors}
\end{figure}

While the fits with both prescriptions look good and do not raise suspicion, the same is not true of the parameter recovery. In Fig.~\ref{fig:posteriors}, we show violin plots of the posteriors of the estimated ejecta masses and velocities of the two components, with the fit with the updated heating-rate prescription in blue and the older prescription in orange. The black crosses indicate the true input parameters. As illustrated, the incorrect prescription, while providing good fits by eye to the observations, fail to recover the true input parameters; estimating an ejecta mass of ${0.003}_{-0.0006}^{+0.0004}$ and ${0.007}\pm0.002~M_{\odot}$ ($68\%$ credible interval) for the two components respectively, fully excluding the true input parameters of $0.015$ and $0.03~M_{\odot}$, respectively. 
Meanwhile, the velocities, while more consistent, also miss the true input velocities, and are more unconstrained by the older prescription relative to the updated heating-rate prescription, with the first component's $68\%$ credible-interval width on the velocity of $18\%$ relative to $4\%$ with the updated prescription. 
The unconstrained posterior is a consequence of the older prescription providing no diagnostic power towards the velocity from the shape of the lightcurve, especially for the high electron fraction first component. 

This bias in estimated ejecta masses and unconstrained velocities has dramatic consequences on the interpretation of kilonovae properties. For real observations where we do not know the true `input' parameters, this raises the need to verify estimated parameters with physical intuition based on order of magnitude estimates alongside other forms of data, such as the spectrum. 
It also raises concerns towards the use of kilonovae alongside gravitational-wave data for constraining the equation of state~\citeg{Coughlin2019, Nicholl2021} and to infer population properties of kilonovae. In particular, when combining events, as the bias in hierarchical analyses grows to the power of the number of events~\citeg{Mandel2019}.

% \begin{figure}[t]
%     \centering
%     \includegraphics[width=0.9\columnwidth]{perturbed_post.png}
%     \caption{Recovered posteriors for salient parameters from a fit with no underlying error on the heating-rate (orange) and one with a $10\%$ uncertainty (blue). The black lines indicate the true input of the simulated data.}
%     \label{fig:perturbedposteriors}
% \end{figure}

%%%%%%%%%%%%%%%%%%%%%%%%%%%%
\vspace{1cm}
\section{Uncertainty in heating rate prescriptions}\label{sec:implications}
The updated heating-rate prescription described in~\citet{rosswog24a}, is a significant improvement over the prescription previously used~\citep{korobkin12a}. However, there are still underlying uncertainties in the nuclear network, reaction rates, nuclear mass model, alongside the thermalization itself that imply some overall uncertainty in the effective heating rate. 

For example, for low electron fractions, $Y_{e} \lesssim 0.2$, the choice between the Duflo-Zuker~\citep{duflo95} and the Finite Range Droplet Model (FRDM) mass formulae~\citep{moeller95} produce a difference of up to an order of magnitude in the heating rate~\citep{barnes16a, rosswog17a}. A similar deviation is also seen with the Gogny-Hartree-Fock-Bogoliubov (HFB)~\citep{Goriely2009} mass model~\citep{Zhu2021, Barnes2021}. 

To investigate the effect of this uncertainty, we simulate synthetic kilonova observations, with both components with low electron fractions of $0.1$ and $0.15$, respectively, consistent with the ejecta of a neutron-star black hole merger. We fit these observations but increase the heating rate for electron fractions $Y_{e} \lesssim 0.2$ by a factor of $10$, with no change to the heating rate for other electron fractions. All input parameters apart from the electrons fractions are kept the same as the simulation in Sec.~\ref{sec:misspecify}. 
The impact of a discrepancy in heating rate up to a factor of $10$ is severe. When fitting with $Y_{e}$ (or opacity) fixed, as commonly done in analyses for AT2017gfo~\citeg{Coughlin2019,Nicholl2021}, we fail to recover the correct input masses and velocities, with no posterior overlap with the true input. In particular, as the heating rate is assumed to be larger (i.e., more energy released per gram of material), we infer significantly smaller ejecta masses by up to a factor of 10. The large difference in heating rate also translates into a visually discernible and significantly worse quality of fit, partly as even with broad priors, there is no suitable combination of ejecta mass, velocity, and opacity compatible with the peak time of the simulated data.
However, if we perform our analysis and keep the electron fractions as free parameters when fitting with a broad prior (we use a uniform prior from $Y_{e} = 0.1-0.5$), we infer broad posteriors on ejecta masses consistent with the true input, but ejecta velocities and electron fraction posteriors with biases significantly larger than the statistical uncertainty. We note that even when fitting with electron fractions kept free, our fits are worse than with the correct input heating rate but not at a level discernible without robust statistical measures.

For studies where electron fractions could be fixed, due to e.g., spectral information, the large difference in fit quality suggest that data alone could distinguish between the heating-rate predictions for the different mass formulae. We note that if the effect of the mass formula is smaller, e.g., up to a factor of $\approx 3$, then photometric data alone is not sufficient to distinguish between the two mass formulae, with no discernible statistical difference in the quality of the fit. 
Analyses where electron fractions and therefore opacities can not be reliably constrained from e.g., the spectrum, and must be kept free when fitting, there is no discernible difference in fit quality. In such scenarios the data will lead to imprecise and inaccurate inferences on ejecta properties.
In general, we expect distinguishing through data is likely to be difficult in practice due to weak constraints on the electron fraction from the spectrum as well as obfuscation from other modelling extensions or the effects of other kilonova modelling systematics discussed in Sec.~\ref{sec:models}.
The discrepancies in inferred properties highlighted above are concerning, as if the discrepancy in the heating-rate due to the different mass formulae is not distinguishable through data alone, then inferred estimates of ejecta properties for lanthanide-rich kilonova ejecta are likely systematically biased by $\gtrsim 50\%$ for the ejecta velocities, while for ejecta masses the statistical uncertainties are underestimating the true uncertainty by up to an order of magnitude. Such uncertainties are far larger than the precision required for any precision constraints on the chemical enrichment of the Universe, or multi-messenger constraints on the equation of state or Hubble constant, while the bias on ejecta velocity, is even more problematic for accurate inferences due to nature of how bias grows when combining events in hierarchical analyses~\citep{Mandel2019}.

Other sources of uncertainty in the effective heating rate, such as due to the nuclear network, reaction rates and thermalization are harder to quantify. However, given the range of estimates from multiple groups, we can conservatively estimate the uncertainty in the effective heating rate to be at least an order of magnitude the~\citep{barnes16a, hotokezaka20, Zhu2021, Barnes2021}. This suggests that the uncertainties above are likely also valid for lanthanide-poor ejecta, i.e., the systematic uncertainty on the ejecta masses could be as high as an order of magnitude, while the ejecta velocities may also be systematically biased.
%%%%%%%%%%%%%%%%%%%%%%%%%%%%
\section{Conclusions}\label{sec:discussion}
In this \emph{Letter}, we have highlighted the dangers of using kilonova heating-rate prescription that are based on a very narrow $Y_e$ range such as the one from~\citet{korobkin12a}. Although this prescription has been widely used as a key ingredient for estimating kilonova properties and designing survey strategies, it is unsuitable for electron fractions significantly different from $Y_{e} = 0.04$, with substantial deviations in output for high electron fractions and at late times. Here, we have shown the differences in individual lightcurves and populations from an updated prescription following~\cite{rosswog24a}. We make these lightcurve models and heating-rate prescription publicly available in open-source package, \program{Redback}~\citep{sarin23_redback}\footnote{\program{Redback} is available at \url{https://github.com/nikhil-sarin/redback}}. 

We have also shown the dangers of using the incorrect prescription by fitting realistic, simulated observations with the Vera Rubin Observatory. We find that while fits look convincing by eye and do not raise suspicion, they fail to recover the correct input ejecta properties with biases greater than the estimated statistical uncertainty. 
This has severe consequences for judging the relevance of neutron star mergers for the chemical enrichment of the cosmos, the interpretation of kilonovae and their use in multi-messenger studies to constrain the equation of state or Hubble constant. In a real observation, where we do not know the true `input' parameters, this highlights the need to scrutinize estimated parameters with physical intuition and consistency with different theoretical studies due to the limitations inherent in detailed modelling of kilonovae. 

Our study demonstrates that given current estimates of the uncertainty in effective or `naked' heating rates, we can expect significant biases in the inferred ejecta properties. In particular, we find that analyses where the opacities (or electron fractions) are fixed fail to recover any input parameter, with biases significantly larger than the statistical uncertainty expected from near future observations. For analyses where electron fractions are not fixed, the ejecta masses are poorly measured but consistent with input parameters. However, the ejecta velocities and opacities in this scenario are significantly biased. Such imprecise mass measurements alongside biased opacities (or electron fractions) and velocities may pose significant challenges for correct interpretation of the data.

% limits or significantly biases the set various minimums on the minimum translates into a systematic uncertainty of \red{$\approx 5$ and $\approx 10\%$} on the ejecta masses and velocity, respectively, for lanthanide-poor ejecta. For lanthanide-rich ejecta, the systematic uncertainty is larger up to $\approx 50\%$ due to uncertainties in the nuclear mass formulae. 
We make a number of simplifying assumptions in our analysis in terms of models, including the treatment of thermalization efficiency and simplified relation of electron fraction to opacity. Moreover, our analysis was limited to a small subset of simulated scenarios. However, our limited study highlights the significant systematic uncertainties in inferring kilonovae properties from observations.
This systematic uncertainty limits the precision of constraints on equation of state and Hubble constant from kilonovae, and should ideally be marginalised over when fitting observations such as AT2017gfo, or as additional components in gamma-ray burst afterglows, e.g., GRB 211211A and GRB230307A to ensure that the interpretation of these events are not affected. 
Moreover, this raises the need for caution regarding precision constraints based on kilonovae properties and interpretations of their role in the chemical enrichment of the Universe. 

In this work, we have also introduced and showcased a framework and tool to better understand the impact of different modelling assumptions on inferences of parameters. In the future, we will extend this to work to capture different simulated scenarios and impacts of various other kilonovae modelling uncertainties and assumptions.
%%%%%%%%%%%%%%%%%%%%%%%%%%%%%%%%%%%%%%%%%%%%%%%%%%%%%%%%%%%%%%%%%%%%%%%%%%%%%%%%
\section{Acknowledgments} 
We are grateful to the anonymous referee for a detailed report with many helpful comments. We want to further thank 
Oleg Korobkin for his comments on 
an earlier version.  
N.S. is supported by a Nordita Fellowship. Nordita is supported in part by NordForsk. N.S. and S.R. acknowledge support from the Knut and Alice
Wallenberg foundation through the “Gravity Meets Light” project (PIs: Rosswog \& Jerkstrand).
S.R. has been supported by the Swedish Research Council (VR) under 
grant number 2020-05044, by the research environment grant
``Gravitational Radiation and Electromagnetic Astrophysical
Transients'' (GREAT) funded by the Swedish Research Council (VR) 
under Dnr 2016-06012, by the Knut and Alice Wallenberg Foundation
under grant Dnr. KAW 2019.0112,   by the Deutsche 
Forschungsgemeinschaft (DFG, German Research Foundation) under 
Germany's Excellence Strategy - EXC 2121 ``Quantum Universe'' - 390833306 
and by the European Research Council (ERC) Advanced 
Grant INSPIRATION under the European Union's Horizon 2020 research 
and innovation programme (Grant agreement No. 101053985).
%%%%%%%%%%%%%%%%%%%%%%%%%%%%%%%%%%%%%%%%%%%%%%%%%%%%%%%%%%%%%%%%%%%%%%%%%%%%%%%%
\bibliographystyle{aasjournal} 
\bibliography{paper}
\end{document}